# Evaluation of a User Authentication Schema Using Behavioral Biometrics and Machine Learning

Laura Pryor[1], Jacob Mallet[1], Rushit Dave[1], Naeem Seliya[1], Mounika Vanamala[1] & Evelyn Sowells-Boone[2]

[1] Department of Computer Science, University of Wisconsin – Eau Claire, Eau Claire, WI, USA

[2] Department of Computer Science, North Carolina Agricultural & Technical State University, Greensboro, North Carolina, USA

Correspondence: Rushit Dave, Department of Computer Science, University of Wisconsin – Eau Claire, Eau Claire, WI, 54701, United States of America. Tel: 1-715-836-2490  E-mail: daver@uwec.edu

# Evaluation of a User Authentication Schema Using Behavioral Biometrics and Machine Learning

**Abstract**

The amount of secure data being stored on mobile devices has grown immensely in recent years. However, the security measures protecting this data have stayed static, with few improvements being done to the vulnerabilities of current authentication methods such as physiological biometrics or passwords. Instead of these methods, behavioral biometrics has recently been researched as a solution to these vulnerable authentication methods. In this study, we aim to contribute to the research being done on behavioral biometrics by creating and evaluating a user authentication scheme using behavioral biometrics. The behavioral biometrics used in this study include touch dynamics and phone movement, and we evaluate the performance of different single-modal and multi-modal combinations of the two biometrics. Using two publicly available datasets - BioIdent and Hand Movement Orientation and Grasp (H-MOG), this study uses seven common machine learning algorithms to evaluate performance. The algorithms used in the evaluation include Random Forest, Support Vector Machine, K-Nearest Neighbor, Naive Bayes, Logistic Regression, Multilayer Perceptron, and Long Short-Term Memory Recurrent Neural Networks, with accuracy rates reaching as high as 86%.

## 1. Introduction

The use of mobile devices has become an integral part of our everyday lives and because of this storing personal information on our devices has become commonplace. People now use their devices to hold passwords, store credit card numbers, perform mobile banking, and a myriad of other things involving personal information (Ackerson at al., 2021, Incel et al., 2021, Strecker et al., 2021, Gunn et al., 2019, Mason et al., 2020). However, despite having all of this personal data on devices, the current security measures taken to protect this data are weak. The use of a one-time password or physiological biometric much like the scanning of a fingerprint which has become common in the past years are not as strong as they appear (Alqarni et al., 2020, Siddiqui et al., 2022, Shelton et al., 2018). While both of these authentication methods can be extremely convenient for a user, the convenience of logging into a device does not outweigh the high threat of an attack that these methods carry with them.

Attacks such as eavesdropping or simply the stealing of a password make one-time passwords incredibly easy to attack, and a user's physiological biometrics can often be imitated by attackers. This has led many researchers to look for a better option than the two current methods, and many have suggested the use of behavioral biometrics. Similar to physiological biometrics, behavioral biometrics use specific features from a user to make an authentication decision. However, unlike physiological biometrics which uses a physical trait, behavioral biometrics uses the behavior of a user to make its decision. Since a user's behaviors are made up of such minute details, it is extremely hard for attackers to be able to perfectly imitate a genuine user's behaviors (Siddiqui et al., 2021). Also, behavioral biometrics are not nearly as intrusive as physiological biometrics as they do not have to capture specific data about a user's physical features. Behavioral biometrics can also be used in conjunction with other authentication methods such as passwords to add an extra level of protection. Finally, behavioral biometrics can be used implicitly throughout the entire use of the mobile device, rather than only once at the beginning of using the device like with passwords and physiological biometrics.

In this paper, we propose a new behavioral biometric authentication scheme and evaluate its performance using seven different machine learning algorithms. Machine Learning algorithm's effect on cybersecurity has been positively received in many different research articles (Dave et al., 2022, Strecker et al., 2021, Krishnamoorthy et al., 2018, Siddiqui et al., 2021, Strecker et al., 2021). The algorithms tested include Random Forest (RF), Support Vector Machine (SVM), K-Nearest Neighbor (KNN), Naive Bayes (NB), Logistic Regression (LR), Multilayer Perceptron (MLP), and Long Short-Term Memory Recurrent Neural Network (LSTM-RNN), as they all have been found to perform well with the biometrics we are testing [ML SURVEY]. Out model also uses touch dynamics and phone movement data taken from a fused dataset from two publicly available datasets, as these two biometrics have been found to be effective for user authentication (Acien et al., 2020, Meng et al., 2013, Antal et al., 2015).In this paper, we will develop and evaluate a new user authentication schema using machine learning algorithms such as RF, KNN, SVM, NB, LR, MLP, and LSTM-RNN  and touch dynamics and phone movement to identify genuine users using both a single modal and a multi-modal model.

*1.1 Related Works*

In this section, we will give a general overview of the most recent work done on this topic. There have been several surveys (Meng et al., 2015, Teh et al., 2016, Siddiqui et al., 2021, Alzubaidi & Kalita, 2016) that have been conducted to examine the effectiveness and usability of touch dynamics and phone movement, and all have come to the same general conclusion. The use of touch dynamics and phone movement has the potential to strengthen mobile devices. However, not all schemas are the exact same. While all schemas are evaluated on their accuracy, some researchers increase their focus on usability (Sun et al., 2014, Bo et al., 2013, Ellavarason et al., 2020), others on the evaluation of cost (Meng et al., 2014), or use of continuous authentication (Xu et al., 2014, Buriro et al., 2021, Zhao et al., 2013). Therefore, it is important to acknowledge the most current research in the field to compare. Analyses of the most current schemas are below:

Authors in (Zhang et al., 2021), examined the use of phone movement dynamics as the sole biometric used in an authentication schema. Using the way a user motioned with their hand and the posture in which they held their device, the authors created a schema that used both dynamic and static iteration respectively to make an authentication decision. Three different machine learning algorithms were tested: K-Nearest Neighbor, Support Vector Machine, and finally Random Forest. When comparing the accuracy of the three algorithms, it was found that RF outperformed the other two algorithms as it consistently had Equal Error Rate values below 0.1. Thus, it was used in the training and testing of the model. While all features performed with high accuracies, the fusion of hand motion and hand posture features was an average of 95% with a top accuracy of 99%, compared with the top accuracy of just hand motion which was around 84%, and the top accuracy of hand posture which was around 95% (Zhang et al., 2021). When comparing F1 scores it was confirmed that the fusion of the features achieved the best performance. The authors also looked into the effectiveness of this model against attacks and found that in the first ten attacks on the system, none were successful, and even when the number of attacks reached 100, there was only a maximum of four successes per attacker. Therefore, this proves that the use of behavioral biometrics, specifically, phone movement dynamics, can be effective in not only authenticating users but also keeping their devices safe from attackers. However, the use of a small dataset may affect the actual effectiveness of this schema.

Stylios et al. (2022) propose a schema for continuous authentication using keystroke dynamics and a new data collection tool. Muti-level Perceptron is the only algorithm being used in the training and testing of this model, in an attempt to see if the use of the new collection tool increases the performance of MLP. The new data collection tool proposed is an application that collects keystroke dynamics from the use of a keyboard in the application and sends that information directly to an online database. 39 users participated in data collection which consisted of either memorizing a sentence or number sequence and inputting it or inputting the sentence or number sequence immediately. During testing, the MLP algorithm was able to achieve an accuracy of 97.18% and an EER of 0.02%, (Stylios et al., 2022) which greatly outperformed the performance of MLP in other related works. However, this model did not test against any known attacks, which could influence the effectiveness of the schema.

Li et al. (2021) focused on developing an authentication schema using a double-click action. This schema uses both behavioral and knowledge-based dynamics, and the user needs to not only perform a double-click action, but they also have to double-click the correct location on a pre-selected background image for their password. In this study five different algorithms were tested, Decision Tree, Support Vector Machine, K-Nearest Neighbor, and finally Back Propagation Neural Network. Out of the five algorithms, SVM was found to achieve the best performance as it got an average error rate of 3.8%, and the other algorithms all had AERs of over 7% (Li et al., 2021). In the study, it was found that in the login phase, the schema obtained an accuracy of 95.7% and when participants came back three days later for the retention phase, the model still had a success rate of 93.4% (Li et al., 2021). These results were compared with the results of a related authentication schema, and DCUS either outperformed or had similar results to the other authentication schema. When tested for usability DCUS consistently scored higher in usability than the other schema. Some participants, however, thought that the use of only a single double-click may be insecure, so the researchers mentioned possibly combining DCUS with other schemas such as SwipeVLock (Li et al., 2019) to add in the use of swiping data to better secure the mobile device.

## 2. Method

*2.1 System Overview*

As seen in Figure 1, our framework consists of multiple different phases. The first phase is the fusion of the two datasets. In this phase, we are taking in the data from two datasets that are publicly available. These datasets are BioIdent and H-MOG and are discussed further in our *dataset description* section. In this phase, the raw data from both datasets are fused into one large dataset and then split again into individual datasets of 100 data points for each user. Each user's dataset is then split into training and testing datasets using an 80/20 train-test split. Therefore, for each user we take 80 genuine samples and then 80 impostor samples from the remaining users, ensuring that each of the other users has at least one sample in the training dataset. The remaining 20 genuine samples are then put into the testing dataset along with an impostor sample from each of the other users. After these fused datasets are created, both the training and testing datasets are pre-processed, and feature selection is done. As we will mention in the next section, our dataset consists of 25 chosen features. After pre-processing and feature selection are done, the two datasets split into their training and testing sections. In training, the dataset is run through each of the eight classifiers, and the trained dataset is then stored in the trained data database. After all the training is complete, the testing begins, and the testing data is run through each of the classifiers which have been trained with the datasets in the trained data database. After each classifier finishes, the classifier outputs its authentication decision leaving us with four different authentication decisions for each of the classifiers. The framework for an actual real-world schema would include only the best classifier found in our results, but for testing purposes, the framework includes all seven classifiers.

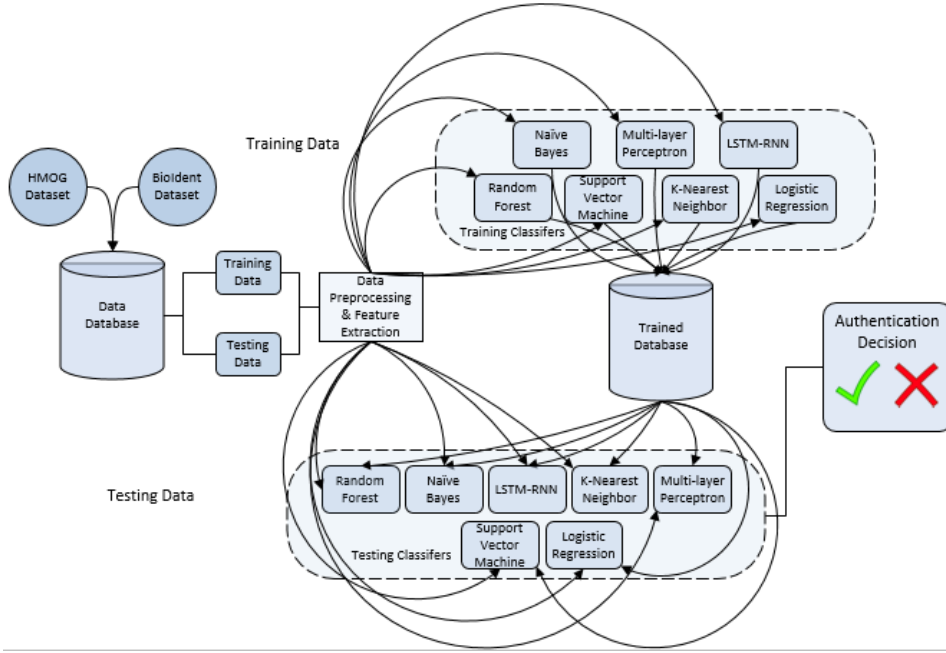

Figure 1. Framework of our model

*2.2 Dataset Description*

For our fused dataset, we used the raw data from two publicly available datasets: BioIdent and H-MOG. The BioIdent dataset was collected by Antal et al. (Antal et al., 2015). and the H-MOG dataset was collected by Yeng et al. (Yang et al., 2014). From the BioIdent dataset, we extracted all of their touch dynamic-based features, and from the H-MOG dataset, we extracted only the features from the sensors. We extracted only the sensors' features from the H-MOG dataset since the BioIdent dataset already included an extremely large and robust dataset of touch features. Both the H-MOG and BioIdent datasets included data from 100 users in multiple sessions, however for our model we tested only one session of 100 samples from 51 users. Thus, after the fusion of the two datasets we had a combined total of 25 features: *user_id, stroke_duration, start_x, start_y, stop_x, stop_y, direct_end_to_end_distance, mean_resultant_length, up_down_left_right, direction_of_end_to_end_line, largest_deviation_from_end_to_end, average_direction, length_of_trajectory, average_velocity, mid_stroke_pressure, mid_stroke_area_covered, acc_x, acc_y, acc_z, gyro_x, gyro_y, gyro_z, mag_x, mag_y,* and *mag_z*. Features from the BioIdent dataset make up the first 16 features in our dataset, and the last nine were from the H-MOG dataset. *User_id* acts as our target for our classifiers and is simply a random number corresponding to a specific user. The feature *stroke_duration* is the amount of time taken to complete a stroke movement. The set of features *start_x, start_y, stop_x,* and *stop_y* refers to the x and y-coordinates at the start and end of the specific movement. *Direct_end_to_end_distance* is the direct length between the start and stop coordinates of the movement. The *mean_resultant_length* is the measurement of the curve that is in the stroke movement. The feature *up_down_right_left* finds the stroke's orientation using displacement. The *average_direction* is the stroke trajectory's average slope. The *length_of_trajectory* is the stroke's overall length. *Average_velocity* is the average velocity of the stroke. The *mid_stroke_pressure* is the pressure collected at the midpoint of the stroke. The *mid_stroke_area* is the covered area by the finger at the midpoint of the stroke (Antal et al., 2015). Finally, *acc_x, acc_y, acc_z, gyro_x, gyro_y, gyro_z, mag_x, mag_y,* and *mag_z*, are the x, y, and z-coordinates for the accelerometer, gyroscope, and magnetometer respectively (Yang et al., 2014). In the end, the dataset consisted of around 5,100 data samples. As mentioned before, the dataset was split into two datasets for training and testing for each user. Each user-specific training set included 80 genuine samples and 80 impostor samples taken from the other 50 users. While some impostor users had more than one sample taken from them, this was not enough to create any significant imbalance in the dataset since only one or two extra samples were taken from the specific impostor. Again, for testing the remaining 20 genuine samples were used along with an impostor sample from each of the 50 remaining users.

### 2.3 Classifiers

The performance of our model was tested using seven different classifiers: Logistic Regression (LR), Long Short-Term Memory Recurrent Neural Network (LSTM-RNN), Multilayer Perceptron (MLP), Random Forest (RF), Support Vector Machine (SVM), K-Nearest Neighbor (KNN), and Naive Bayes (NB). In this section, we will give a quick overview of each of the eight classifiers tested in this study. The pseudocode for how we evaluated each of these algorithms can be seen in Algorithm 1.

```
ALGORITHM 1: Pseudocode for Classifier Evaluations

algorithms = [RF, SVM, KNN, NB, LR, MLP, LSTM-RNN]
metrics = [accuracy, precision, recall, f1Score, eer]

obtain userDatasets

for each user in userDatasets:
    split training and testing data 30/70
    for each algorithm in algorithms:
        create algorithmPredictions list
        add algorithm to algorithmPredictions
        if algorithm = LSTM-RNN:
            scale data
        else:
            continue
        create model using algorithm
        train model using trainingData
        predict using model and testingData
        for each metric in metrics:
            calculate metric
            add metric to algorithmPredictions
        end for
    end for
    end for
    write algorithmPredictions to file
end for
```

### 2.3.1 Logistic Regression

Beginning with LR, Logistic Regression is a supervised learning technique that is used heavily in classification models. The basis of LR is that it uses the sigmoid function to map the probabilities to predictions. The sigmoid function creates an S-shaped curve between the numbers 0 and 1, and any real value can be put in this curve between 0 and 1. Logistic Regression uses a threshold in order to make its class label decision, in our model we used the default threshold of 0.5, so anything above 0.5 would be classified as a genuine user and anything below 0.5 would be classified as an impostor. Since our model has multiple class labels, our logistic regression model outputs the predicted class label for the user rather than the binary classification of 0 or 1. We then take these class labels and assign them binary classifications based on if the user is what is expected, which would then be granted a classification of 1, or the user was not the expected user, which would then be granted a classification of 0.

### 2.3.2 Long Short-Term Memory Recurrent Neural Network

LSTM-RNN is a deep learning method which uses LSTM networks to make predictions. LSTM networks are a type of recurrent neural network, therefore this algorithm is considered a deep learning algorithm. RNNs work in a way where they can remember past information by using recurrent units and then use that information when working with new data, however unlike humans, RNN cannot remember information long term. LSTM works similarly to RNNs but the processes done inside of the recurrent unites differ, as LSTM uses a cell state to act as a long-term memory. LSTM uses a forget gate to control whether information from the past timestep should be forgotten or not, and it also uses an input gate to calculate the importance for the information being gained in the new timestep. These gates allow for the current timestep to be updated with only the most important information needed to make an accurate decision. Therefore, instead of all of the information from the past timesteps being forgotten over time as new timesteps erase the past information, the current timestep includes only the most important information from all of the past and current timesteps. Which in turn, improves learning over using just a basic RNN. For our loss function we used binary cross-enrtopy since our outcomes were binary.

### 2.3.3 Multilayer Perceptron

MLP is another deep learning algorithm that uses artificial neural networks. MLP uses a perceptron that includes multiple different input, output, and hidden layers to create a class label. A perceptron was created to mimic the human function of perception, in which the perceptron is made up of neurons that use a weighted sum and an activation function to make an output. MLP is a feed-forward algorithm which means that each layer that makes up MLP is connected to the one before and after it and each layer is a row of neurons. In the case of our model, since we are using multiple class labels our MLP uses a softmax activation function to output the different class labels from the output layer. Then once all of the labels have been outputted, we transform those labels into binary classifications based on if they are a genuine user or not.

### 2.3.4 Random Forest

RF is a machine learning algorithm that uses a "forest" of multiple different decision trees to create class labels. For each data point that is put through our model, RF takes this data point and runs it through each of the predetermined amount of decision trees that were created. Each tree would then make its own class label for the data, and then all of the predictions that were created by the trees are then "voted" on to see which class is most common, which is then outputted as the final result. In order to add an extra level of analysis than just what a basic decision tree algorithm has, each decision tree in the forest is slightly different than all the others which increase the amount of analysis done on the data points. Since we have multiple classes in our model, each decision tree outputs what they predict the class of the data point to be, and then after each final prediction is voted on we turn that into a binary classification similar to what we did with LR.

### 2.3.5 Support Vector Machine

SVM is used very often in authentication schemas using touch dynamics and phone movement and uses hyperplanes to create its predictions. In SVM, an n-dimensional space is created based on the number of features, in which multiple hyperplanes reside. Each of the hyperplanes holds all of the classified data points, however, this algorithm does not use all of the hyperplanes. SVM's goal is to find the hyperplane that has the maximum distance between support vectors. Support vectors are the data points of each distinct data group that lies the furthest out from the group. Therefore, SVM is trying to pick the hyperplane that has the maximum distance between each of the data groups. The reason for trying to find the maximum distance is so that it is easier for the model to predict the targets of different data points because the farther the different groups are away from each other, the more likely it is that the data points that belong to each group will be very distinct to that data grouping. Again, similar to RF and LR in our model SVM predicts a class label instead of a binary prediction, we then transform this class label into a binary prediction similar to how we do it with the other two algorithms.

### 2.3.6 K-Nearest Neighbor

Another machine learning algorithm that is evaluated is KNN. KNN also uses the distance between classified and unclassified points to help make its predictions. In KNN, every time a new data point is added to the database, k neighbors are chosen to "vote" on the class of the new data point. K stands for the predetermined amount of neighbors the schema wants the algorithm to use in its voting. As with the other algorithms, our KNN algorithm predicts the class label of the data point and not a binary classification. Similar to the other algorithms, our binary transformation is done after our schema receives the class label for the data point.

### 2.3.7 Naïve Bayes

The final algorithm we are evaluating is NB. NB is a fast and simple algorithm and unlike many of the other algorithms we are evaluating in this schema, NB uses a probabilistic approach to make its decisions. NB uses the Bayes Theorem which calculates conditional probability. In NB it is thought that none of the features being inputted are related to each other, therefore all features are independent and contribute equally to the probability of, in the case of our model, the label of the user. However, the naivety of this model can also be its downfall, since a completely independent set of predictors of an event is nearly impossible to happen.

*2.4 Success Metrics*

To evaluate the success of the algorithms we are testing in our model, we used the following success metrics: accuracy (ACC), precision, recall, F1 Score, and Equal Error Rate (EER). Beginning with accuracy, ACC is the measurement of what percentage of user samples are correctly identified. Precision, as it says in the name measures how precise our model is, therefore it looks at the rate of true positive classifications out of all positive classifications. Thus precision measures the percentage of correctly classified positive samples. Recall is similar to precision but instead looks at the sensitivity of the model, thus it calculates the percentage of true positives out of all possible positive classifications. F1 Score calcluates accuracy, but only using the values from precision and recall. Therefore F1 Score can measure how false negatives and false positives can effect the model's performance. Finally EER in which the model's False Acceptance Rate and False Rejection Rate are equal. Thus, we want this value to be low to ensure that our model is not producing many false acceptances and rejections.

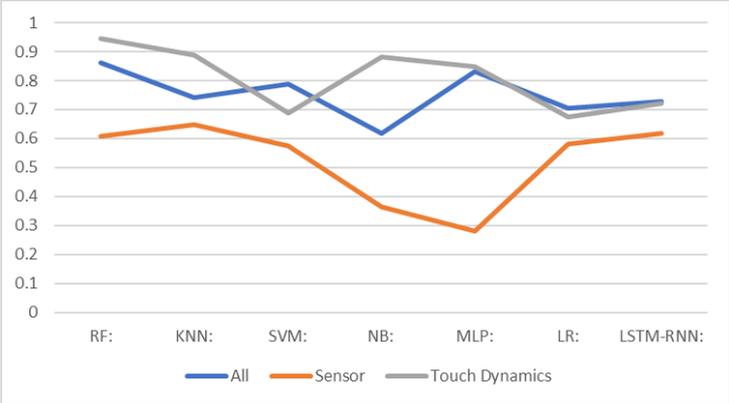

Figure 2. Average accuracies of all users

## 3. Results

To evaluate our model, we examined the performance when inputting features from different combinations of various sensors, resulting in a total of 15 distinct combinations. In our results, we tested the measured results of each success metric using both the full sample of users and a random sampling of 10 users. We evaluated the results of the three main sections: all features, touch dynamics only, and sensors only, using all users, and evaluated the in-depth combinations of features using the random sampling of 10 users.

*3.1 Touch Dynamics*

This subsection only deals with data captured by the touchscreen, which primarily came from the BioIdent

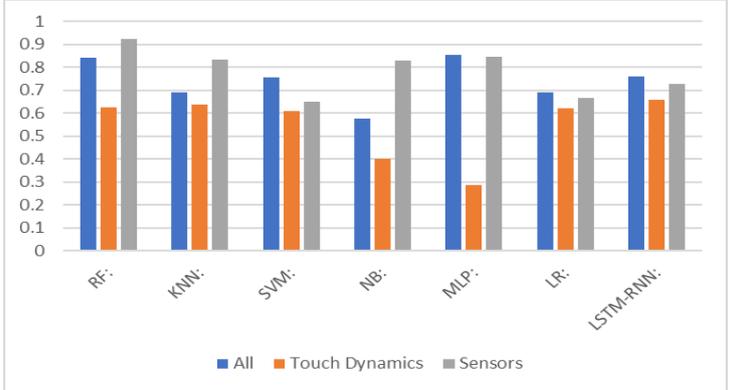

Figure 3a. Average accuracies of 10 random users

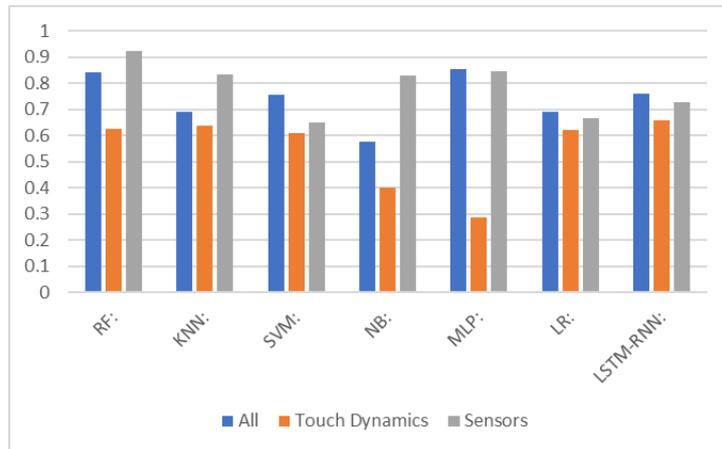

Figure 3b. Average accuracies of 10 random users

dataset. Five out of the seven algorithms produced accuracies that fell in the range of 68.8% to 88.7%, with the highest coming from RF the results of which can be seen in Figure 2. These results differed from the sampling of 10 users, as that sampling gave a slightly lower accuracy range with five of the seven algorithms falling in the range of 60.9% to 65.7% which can be seen in Figure 3a and Figure 3b. One outlier came from MLP which yielded a 28.0% accuracy. Unlike with the large user sample however, LSTM-RNN performed the best in the smaller grouping with K-NN becoming the second-best algorithm. Across the board, there was a wide range of outcomes for our recall metric. This ranged from 61.0%, which came from LSTM-RNN, to 100%, produced by MLP seen in Figure 4. One thing interesting to note is that Naive Bayes yielded one of the worst accuracies, but also one of the highest recall scores. This means that the algorithm has a high True Acceptance Rate (TAR), despite not being able to accurately identify users consistently. Our sample of 10 users also had high recall results which can be seen in Figure 5. Our EER was fairly high for every algorithm we evaluated when given solely touchscreen

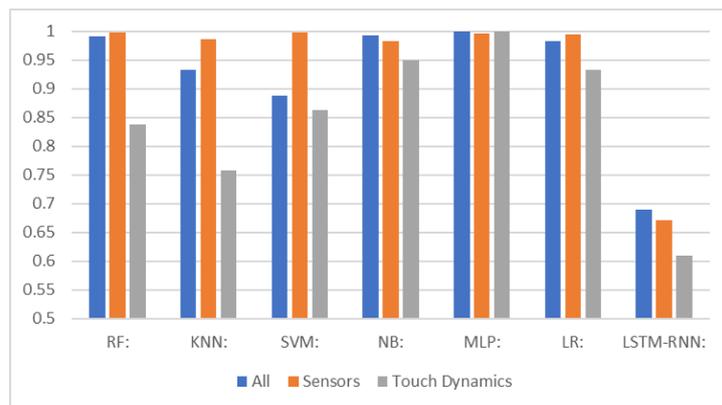

Figure 4. Average recalls of all users

. For the EER average for all users, LR produced the lowest with 31.1%, and both MLP and LSTM-RNN produced the highest EER with 50.0% as seen in Figure 6. The lowest for our sample of 10 users was produced by logistic regression, which was 28.8%. MLP yielded the highest EER, at 50.0% which can be seen in Figure 7. None of our algorithms were very precise when dealing with solely touchscreen features, with this metric ranging from 28.6% to 46.2% for the sample of 10 users seen in Figure 8 and ranging from 47.3% to 84.4% for the sample which included all users which is seen in Figure 9. One takeaway from this section is that our model tends to perform better when there is a larger sample of imposter data to distinguish from, as our results with the larger sample size tends to yield better results.

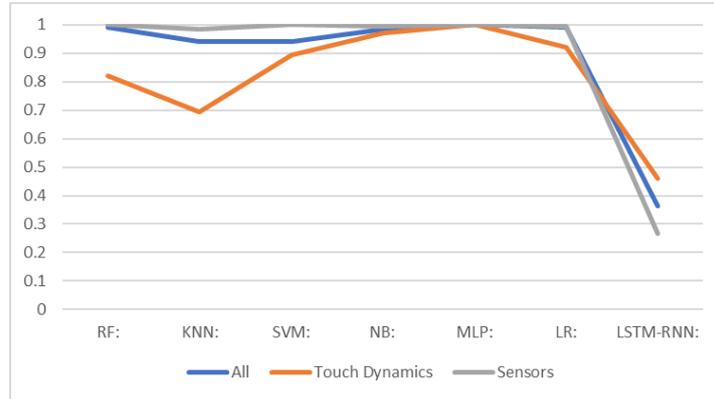

Figure 5. Average recalls of 10 random users

*3.2 Phone Movement*

In this subsection, we describe our results that were yielded when our algorithms were given features that were captured by sensors only, along with various combinations between the magnetometer, gyroscope, and accelerometer. We will first discuss the results from each sensor isolated. The magnetometer yielded high scores for multiple different metrics. For our sample with 10 random users, RF, KNN, MLP, and NB all yielded accuracies above 82.7%, whereas SVM and Logistic Regression were the worst performing algorithms regarding accuracy as seen in Figure 3. Our sample with all users performed similarly with RF, KNN, MLP, and NB all yielding accuracies over 84.7% as seen in Figure 2. Although the same number of users were available in the dataset for a singular sensor, there were only three features being given to the algorithms in this section. This equates to less data as a whole for the algorithm, which could be the reason responsible for SVM having a poor accuracy for sensor only evaluations. Similar to the trend we observed earlier, SVM and Logistic Regression who yielded the lowest accuracies for the sample of 10 users, also provided the best recall score, at 99.9% which can be seen in Figure 5. The results for the sample of all users performed similarly to the smaller group size as seen in Figure 4. For EER, the sample of all users performed slightly better than touch dynamics alone, with most algorithms achieving EERs between 10.7% and 26.12% as seen in Figure 6. RF was an outlier however getting an EER of only 5.4%. The majority of the algorithms achieved an EER lower than 11.1%, with the lowest rate being 3.3%, from KNN seen in Figure 7. The isolation of the magnetometer sensor yielded some of the best results of isolating each sensor, and across all of our evaluations as a whole. The numbers observed as a result of evaluating our model with data from the accelerometer or gyroscope were not able to reach the results the magnetometer data produced. Similar to the magnetometer sensor results, SVM also produced one of the worst accuracies, 41.6% for the accelerometer and 28.6% for the gyrometer, yet a 99.9% recall score for each of the sensors, the accuracies of which can be seen in Figure 10. Results from the gyrometer sensor data showed EER rates rose to around 50% for LSTM-RNN, MLP, SVM, seen in Figure 11, which were some of the highest EER yielded by this for any data combination we evaluated. Regarding the accelerometer results, F1 scores, seen in Figure 12, rose as high as 71.7% from Random Forest, which also yielded the highest F1 score for the gyrometer sensor as well, at 59.2%. This indicates Random Forest provides a good recall and precision balance relative to other algorithms we evaluated. Recall scores had a wide range of outcomes for each sensor, but specifically the gyrometer. Logistic Regression, MLP, Naive Bayes, SVM, KNN, and Random Forest all yielded a recall score greater than 85.0%, while LSTM-RNN was an outlier at 31.5% shown in Figure 13. Overall, recall scores were high besides two outliers, but precision scores were much lower for both sensors, with the highest score achieved by Random Forest at 66.7% using the accelerometer data. In addition to evaluating each sensor, we combined the data from these sensors in four different ways: the accelerometer and gyrometer, the accelerometer and magnetometer, the gyrometer and magnetometer, and finally all three sensors together. Combining the accelerometer and magnetometer yielded some of our best metrics we saw from our model. Accuracies on average reached as high as 97.9% from Random Forest, and 97.6% from Naive Bayes seen in Figure 14.

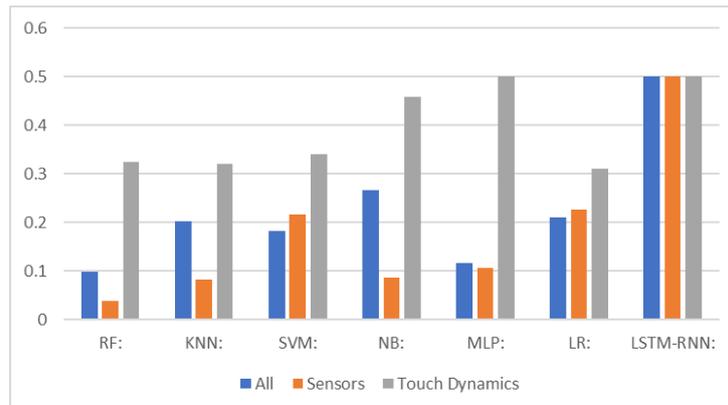

Figure 6. Average EERs for all users

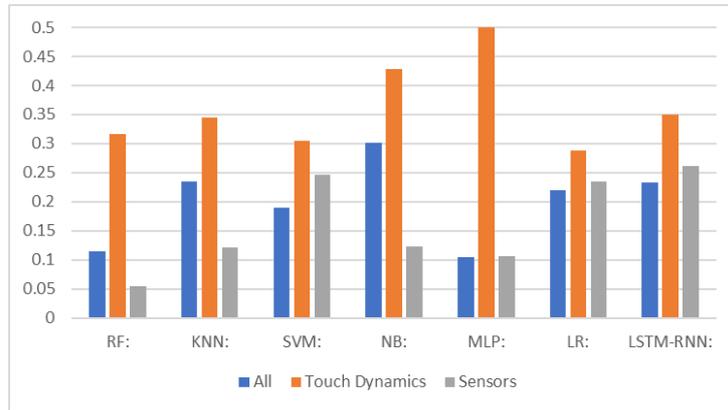

Figure 7. Average EERs for 10 random users

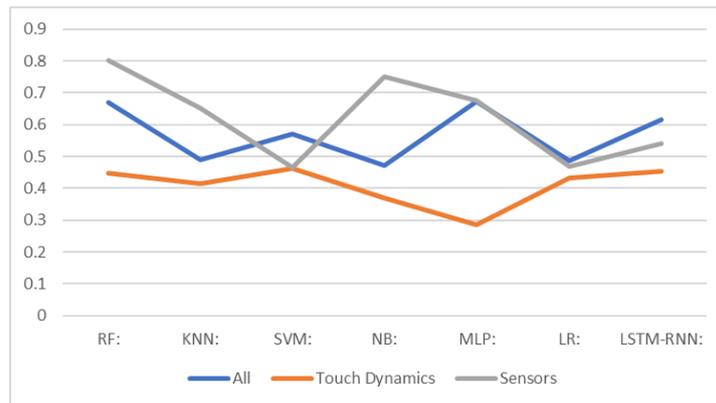

Figure 8. Average precision for 10 random users

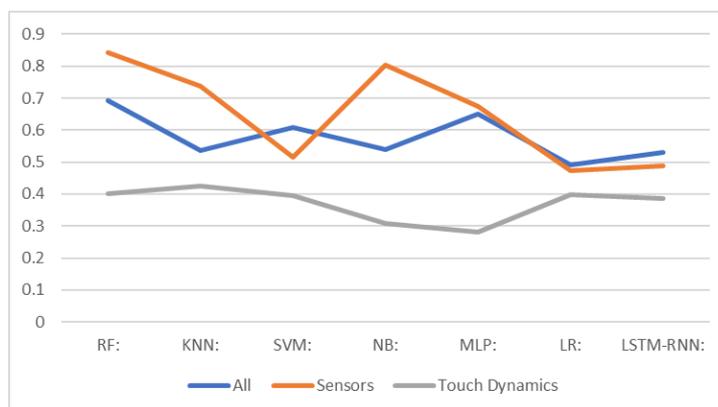

Figure 9. Average precision for all users

Random Forest also produced the best recall, F1 score, and EER for this dataset at 99.9%, 96.4%, and 1.5% respectively. All of these scores were the best regarding each metric we observed for any combination of data we evaluated our model with. The results of which can be seen in Figures 15, 16, and 17 respectively. Other algorithms also performed well regarding accuracy, with LSTM-RNN and MLP also staying above 80%. The lowest accuracy came from Logistic Regression, which yielded a score of 65.4%. Naive Bayes produced thehighest precision score at 99.1%, which was again the highest score for precision across all combinations of data shown in Figure 18. Combining the accelerometer and magnetometer sensor data yielded some of the best results for all five metrics compared to any of the other combinations of sensors we evaluated. The combination

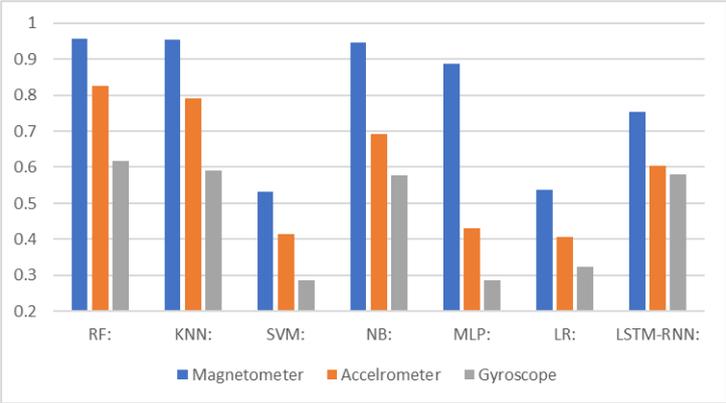

Figure 10. Average accuracies of sensors for 10 random users

that produced some of the lowest results between the sensors was the accelerometer and gyrometer. 75.7% was the highest accuracy achieved which came from Random Forest. Precision was also low, as the highest result came from Random Forest again at 52.8%. A common theme observed throughout our model has been that recall is higher than precision which continued with this sensor combination. Six out of seven of our algorithms had a recall above 81.5%. Combining the data obtained from the gyroscope and magnetometer produced similar, but slightly worse results to the accelerometer and magnetometer combination. The best accuracy, recall, F1, and EER were again achieved by Random Forest, at 96.3%, 99.9%, 94.0%, and 2.6% respectively. Naive Bayes again yielded the best precision score at 90.2%. Overall, this combination did outperform the accelerometer and gyroscope in most metrics, but not quite as high of scores on average as the accelerometer and magnetometer. The common factor between the two best performing combinations is the magnetometer, which could indicate this sensor does well in authentication. This idea is also supported from earlier, when we found the magnetometer performed the best individually out of all three sensors. When combining all three sensors, results were not as well as other combinations of sensors. Random Forest was yet again the best performing algorithm,

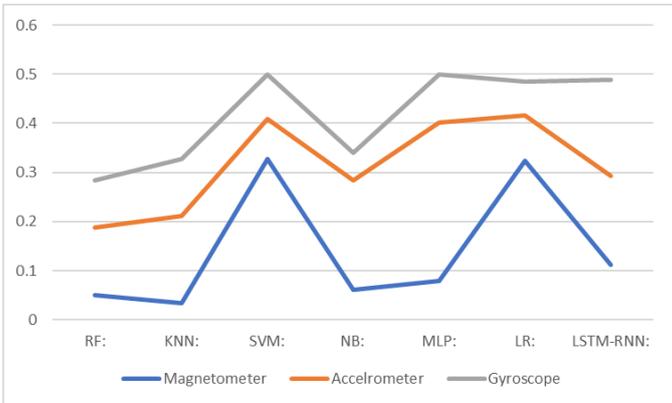

Figure 11. Average EERs of sensors for 10 random users

with the best score for all of the metrics. Random Forest achieved an accuracy of 92.2%, precision of 80.3%, recall of 99.9%, F1 score of 88.7%, and an EER of 5.4%. These metrics were not as high as other combinations of sensors but are still higher than the majority of scores observed from other combinations that involved touchscreen data. Recall scores were very high for most algorithms, with Random Forest, KNN, SVM, Naive

Bayes, and Logistic Regression all achieving a score greater than 98.5%. One major outlier that was observed came from LSTM-RNN, which yielded one of our lowest recall scores throughout the entirety of our evaluations at 26.5%. Since recall is closely related to the F1 score, this metric was also very low for LSTM-RNN at 27.4%. F1 scores didn't reach the heights most of the other combinations achieved, as no algorithm scored higher than 88.7% for this metric.

*3.3 Touch Dynamics & Phone Movement Combined*

This subsection will entail the results regarding varying combinations of touchscreen and sensor data. The results of these combinations can be seen in Figures 14, 15, 16, 17, and 18. First, we evaluated data captured by the touchscreen combined with an individual sensor, meaning three combinations were tested. Touch dynamics and the magnetometer sensor yielded the best scores across any combination of touch dynamic data and sensor. These scores were produced by Random Forest, with an accuracy of 89.1%, precision of 74.2%, recall of 99.9%, F1 score of 84.7%, and an EER of 7.6%. Recall scores were generally high for the combination of these sensors, with 6 out of 8 of the algorithms being at or above 92.5%. As far as accuracy is concerned, there was a fairly high floor with this combination of data, as no algorithm had an accuracy below 74.4%, which was Logistic Regression. This was unlike the results from the touch dynamics and gyrometer combination, which were much

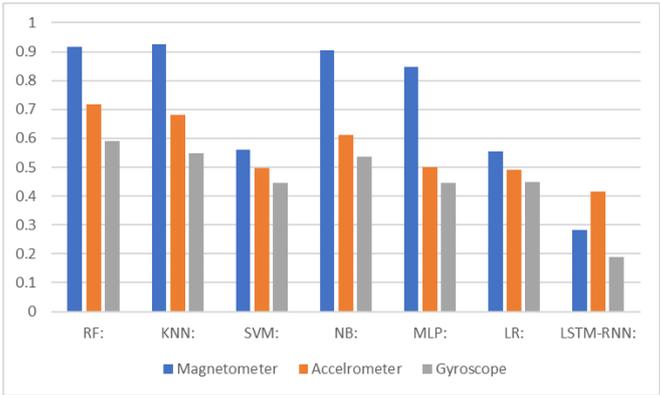

Figure 12. Average F1 Score of sensors for 10 random users

worse for most of the metrics observed. Recall scores, which have been consistently high throughout the model, dropped to a low of 23.5% for LSTM-RNN, with this algorithm also yielding the 2nd highest accuracy for this combination at 66.1%. MLP yielded the lowest accuracy and EER of 28.9% and 49.9% respectively, but a high recall score of 99.9%. Precision as a whole was very low, as the highest score came from Random Forest at 48.9%. Results for the combination of touch dynamics and the accelerometer were slightly higher, but not as high as the combination of touch dynamics and magnetometer. Random Forest yielded the highest accuracy at 76.7%, while EERs ran as high as 44.3% and as low as 16.5%. Recall remained similar to other combinations as

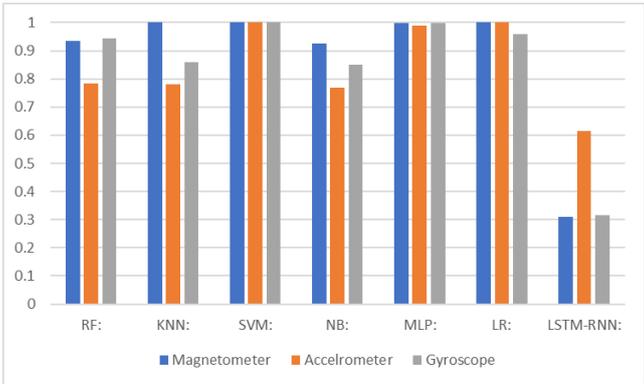

Figure 13. Average recall of sensors for 10 random users

over half of the classifiers scored 94% and above. The touch dynamic data was then combined with 2 different sensors and evaluated. The same pairs of sensors discussed in the previous subsection were now combined with touch dynamic data, resulting in three combinations of sensor input. For the combination of touch dynamics, accelerometer, and magnetometer, Random Forest yielded the best accuracy, precision, F1 score, and EER, at

84.0%, 65.1%, 77.9%, and 12.0% respectively. Naive Bayes produced the highest recall score of 99.9%, but also yielded the worst precision at 46.4%. EER scores remained relatively low for all algorithms, as the highest EER came from Naive Bayes at 29.5%. This was similar to the results from combining touch dynamics, gyrometer, and the magnetometer, where the highest EER was 22.3%, coming from Naive Bayes. Six out of the seven algorithms yielded recall scores at 96.5% or higher. Accuracies observed from this combination of data ranged from 68.9% to 83.7%. One last sensor combination evaluated was between touch dynamics, accelerometer, and gyrometer data. The scores observed here were lower than most combinations. The highest accuracy came from Random Forest at 73.9%, with the lowest coming from MLP at 37.1%. This combination also yielded lower precision scores, with the highest being 54.7% from Random Forest. Finally, when adding all data features together our sample of all users were able to produce the highest accuracy of 86.3% using RF which was similar to the results of our smaller group of 10 users in which MLP, the second-highest algorithm with all users, achieved the highest accuracy with 85.1% as seen in Figures 2 and 3 respectively.

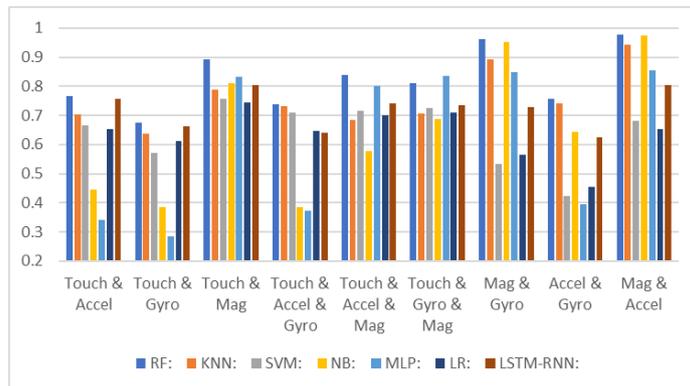

Figure 14. Average accuracies of feature combinations for 10 random users

## 4. Discussion

### 4.1 Discussion & Analysis of Results

One significant finding observed throughout the entirety of our model was consistently high recall averages, which indicates that our model is performing well at classifying the genuine user. In other words, our model has a high True Acceptance Rate (TAR) overall. This is promising, as a common issue that troubles authentication schemes is rejecting the genuine user or having a high False Rejection Rate (FRR). Having a model that performs well in this area is key in order to have good usability. Sufficient usability is vital to improve the user's confidence in the schema. On the other side of things, an authentication scheme that denies the genuine user access frequently is extremely inconvenient for the user. One metric that is related to recall and scored consistently lower, is precision. Precision indicates the proportion of correctly identified genuine users, therefore the lower the precision, the less secure our model is. Authenticating users that are not the genuine user leaves any sensitive information on the device susceptible to attackers. Balancing security with usability is not a unique issue to this schema, as many other authentications have been troubled by the same problem. Oftentimes an attempt to raise precision or recall may lower the other. For example, increasing how selective our model is may increase the rate the genuine user is denied access, resulting in a lower recall for our model. Utilizing a larger dataset to our model could help this issue, as simply providing more examples of what the genuine user's data looks like to our model could improve the security as a whole, by increasing the rate our model rejects imposters. In a real world context, the majority of users would most likely prioritize security over usability. Finding an improved balance between security and usability is essential in order for our authentication schema to be applied in a real-world scenario. Another significant finding seen across all of our evaluations was that Random Forest consistently yielded some of the best metrics on average for the majority of the data combinations we evaluated. This indicates that Random Forest may perform well in authentication, especially when the incoming dataset is limited in size, like it was in our model. As mentioned previously, Random Forest produced the best accuracy, recall, F1, and EER, at 97.9%, 99.9%, 96.4%, and 1.5% respectively, across all evaluations of our data. These numbers came as a result of combining the data captured by the accelerometer and magnetometer sensors. As suggested earlier, our dataset was limited in size, which did affect our results. Expanding this dataset as a whole could result in other algorithms improving their results and possibly surpassing Random Forest. Some algorithms, specifically SVM, don't perform as well with smaller datasets as others, like Random Forest do. Expanding the dataset would help provide clarity on the context of our results regarding each algorithm. Regarding the combinations of data we evaluated, another significant finding from our schema was that results tended to be higher when data from the magnetometer sensor was accessible to the model. After isolating data involving touch dynamics as well as each sensor and evaluating our model, the magnetometer data yielded the best results for most of the metrics in relation to the other individual sensors. Some of these results produced from this sensor individually were close to the highest scores we observed across all of the data combinations evaluated. As a result of evaluating the combination of the magnetometer and accelerometer data, we observed the highest individual scores for all five of the metrics evaluated. There appears to be a correlation between improved results after introducing the magnetometer data to a combination for most algorithms, more so than any other sensor. A common theme like this throughout the entirety of our model could be an indicator that data coming from the magnetometer sensor does well in providing information to allow our model to identify the genuine user more consistently.

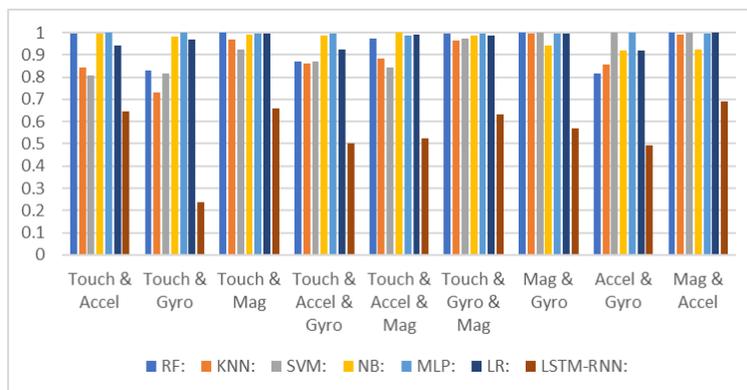

Figure 15. Average recall of feature combinations for 10 random users

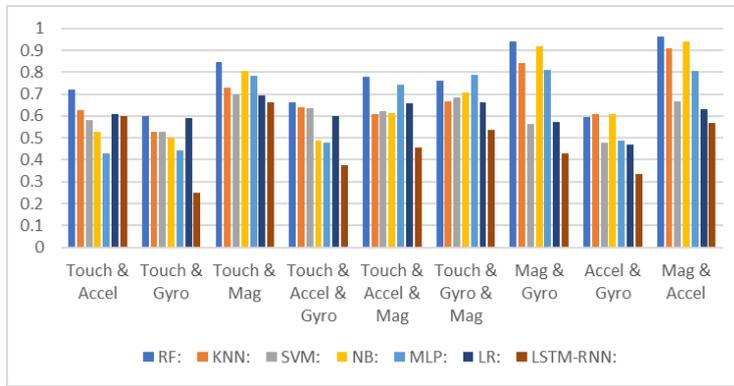

Figure 16. Average F1 Score of feature combinations for 10 random users

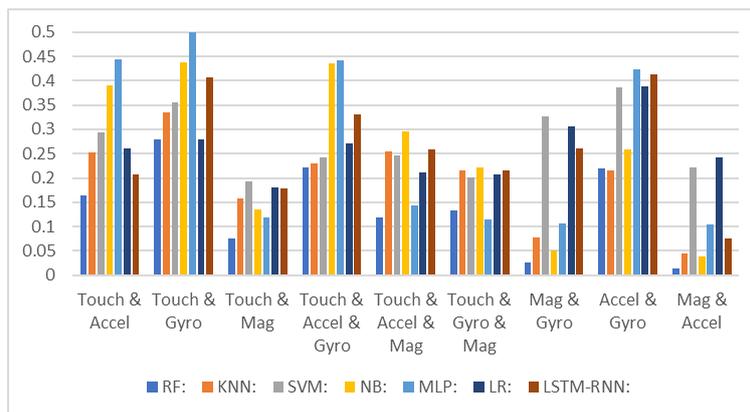

Figure 17. Average EER of feature combinations for 10 random users

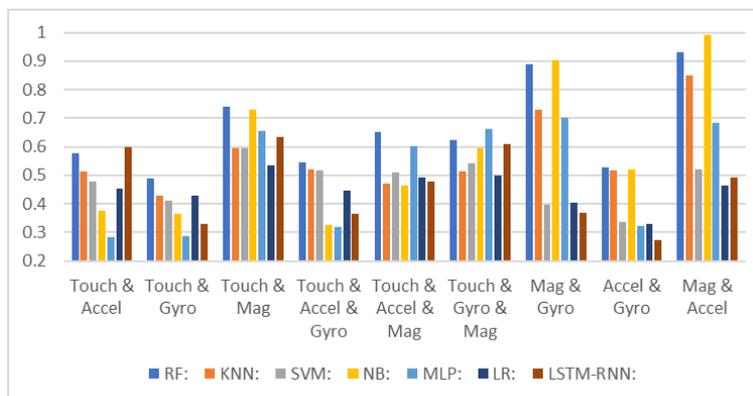

Figure 18. Average precision of feature combination for 10 random users

*4.2 Limitations*

Several different issues still could be affecting our model, despite our authentication scheme showing promise. One variable that influenced our results was our inability to distinguish between single and multi-touch data points. The datasets we obtained from online did not distinguish between these two types of touch interactions. This could result in accuracies seeming higher artificially, meaning the results may seem better than they should be. Another issue discovered throughout this research was the limited size of the dataset. Even after fusing two datasets together, the dataset was still an issue for our model, especially on a user by user basis. A possible issue that could be caused by the limited dataset is our model producing inconsistent results. Inconsistency issues within our model could skew accuracies positively or negatively artificially. Another issue that appeared from fusing two datasets together was that we were unable to make certain that new users being created from the two datasets were being fused with data from two similar users. Users that have little experience interacting with smartphones being in the same dataset as others who are highly knowledgeable in their use of technology could lead to inaccurate data for our newly created users.

*4.3 Conclusion & Future Work*

This research developed a multi-modal authentication model, which utilizes multi-modal behavioral biometrics, specifically touch dynamics and phone movement. We used two publicly available online datasets, namely the BioIdent H-MOG dataset, which captured their data using the touchscreen, accelerometer, magnetometer, and gyroscope sensors. Our model then preprocessed the data and performed feature extraction. After this, several machine learning algorithms were applied, including Random Forest, K-Nearest Neighbors, Support Vector Machine, Naive Bayes, Multi-Layer Perceptron, Logistic Regression, and a Long Short-Term Memory Recurrent Neural Network. After evaluating multiple combinations of touch dynamics and phone movement sensor data, our model was able to achieve an average accuracy from ten random users of 97.9%, which was yielded from Random Forest classifier, using data captured by the accelerometer and magnetometer. As a whole, our results exhibit the potential that behavioral biometrics possess to authenticate users on mobile devices. Despite our results, there is still further work that is required in order for this schema to be implemented on mobile devices in a real world scenario. One major obstacle still in the way of this schema being applied in a real world context is finding the right balance between usability and security. These two requirements are vital in order to take the next step with this model. Currently, our model is too far skewed toward the usability side of the spectrum. As far as future work is concerned, we will carry on with improving this model by looking to create a new, more robust, dataset than ones currently available to the public. While creating this dataset, we will put an emphasis on distinguishing data points between single and multi-touch gestures, as this is a shared issue among research efforts along with our own. The result of this new dataset that distinguishes between single and multitouch gestures could lead to our model achieving higher accuracies and a more consistent representation in totality.

**Acknowledgments**

Funding for this project has been provided by the University of Wisconsin-Eau Claire's Karlgaard Computer Science Scholarship Foundation as well as the University of Wisconsin-Eau Claire's Office of Research and Special Programs Student-Faculty Collaboration Grant.